\documentclass[12pt]{article}
\pdfoutput=1
\usepackage[utf8]{inputenc}
\usepackage{amsmath}
\usepackage{amssymb}
\usepackage{units}
\usepackage{graphicx}
\usepackage{slashed}
\usepackage{array}
\usepackage[leftcaption,raggedright]{sidecap}
\usepackage{caption}
\usepackage[hidelinks]{hyperref}
\usepackage[sorting = none]{biblatex} %Imports biblatex package
\newcommand\pubnumber{NuPhys2023-Hunt-Stokes}
\newcommand\pubdate{\today}

\def\napoli{University of Oxford}

\def\support{\footnote{
  Science and Technologies Facilities Council.
}}

\def\Title#1{\begin{center} {\Large #1 } \end{center}}
\def\Author#1{\begin{center}{ \sc #1} \end{center}}
\def\Address#1{\begin{center}{ \it #1} \end{center}}

\newcommand\pubblock{\rightline{\begin{tabular}{l} \pubnumber\\
         \pubdate  \end{tabular}}}
\newenvironment{Abstract}{\begin{quotation}  }{\end{quotation}}
\newenvironment{Presented}{\begin{quotation} \begin{center} 
             PRESENTED AT\end{center}\bigskip 
      \begin{center}\begin{large}}{\end{large}\end{center} \end{quotation}}

\def\beq{\begin{equation}}
\def\eeq#1{\label{#1}\end{equation}}
\def\eeqn{\end{equation}}

\def\beqa{\begin{eqnarray}}
\def\eeqa#1{\label{#1}\end{eqnarray}}
\def\eeqan{\end{eqnarray}}

\let\bar=\overbar

\def\Dslash{\not{\hbox{\kern-4pt $D$}}}
\def\dslash{\not{\hbox{\kern-2pt $\del$}}}

\def\msb{{\bar{\ssstyle M \kern -1pt S}}}

\begin{document}
\begin{titlepage}
\pubblock

\vfill
\Title{Calibration of the Scintillation Timing in SNO+ using In-Situ Backgrounds}
\vfill
\Author{ Rafael Hunt-Stokes\support}
\Address{\napoli}
\vfill
\begin{Abstract}
The SNO+ Collaboration has recently concluded loading its liquid scintillator with PPO, the primary fluor, and the loading of the wavelength shifter, bisMSB, is ongoing. For each stage of the experiment, reliable position and energy reconstruction is essential, and in the face of a changing scintillator cocktail, methods have been developed to rapidly calibrate the SNO+ optical model. To this end, a novel technique for calibrating the time response of the SNO+ liquid scintillator using in-situ backgrounds was developed. By using in-situ coincident backgrounds, it is possible to calibrate both the $\beta^-$ and $\alpha$ time responses, as well as facilitate frequent monitoring of the background levels without compromising the radiopurity of the detector. Accurate calibration of the liquid scintillator emission times allows the exploration of time-based particle discrimination, such as between single-site ($\beta^-$) and multi-site ($\beta^- \gamma$) interactions.

\end{Abstract}
\vfill
\begin{Presented}
NuPhys2023, Prospects in Neutrino Physics\\
King's College, London, UK,\\ December 18--20, 2023
\end{Presented}
\vfill
\end{titlepage}
\def\thefootnote{\fnsymbol{footnote}}
\setcounter{footnote}{0}

\section{Introduction}
The SNO+ detector is a large liquid scintillator detector located 2 km underground at SNOLAB, the underground research laboratory in Sudbury, Canada \cite{Albanese_2021}. The detector has transitioned through a number of operational phases, characterised by the composition of the deployed liquid scintillator. These phases are: partial-fill (LAB + 0.6 g/L PPO), full-fill (LAB + 2.2 g/L PPO) and the current bisMSB loaded phase (LAB + 2.2 g/L PPO + 2.2 mg/L bisMSB).

As the composition of the scintillator changes, so too does the scintillator’s photon emission time profile (see figure 1). Therefore, each phase requires accurate calibration of the scintillator’s time response in order to continue performing reliable position reconstruction and particle identification. Additionally, due to the stringent radiopurity requirements for various analyses, it is desirable to leave the detector untouched as much as possible. To this end, a novel in-situ calibration technique has been developed using intrinsic $^{238}$U decay chain daughters: Bismuth-Polonium 214 (BiPo214) coincidence events. $^{214}$Bi provides a source of $\beta^-$ interactions and $^{214}$Po provides a source of $\alpha$ interactions.

\section{Scintillation Emission Time Model Tuning}
The shape of an organic scintillator's emission time profile results from the complex interplay of energy transfer processes between the chemical species comprising the scintillator. A non-exhaustive list of the dominant effects includes \cite{birks2013theory}: 
\begin{itemize}
    \item Intrinsic timing of each species
    \item Charge density of interacting particles
    \item Concentrations of primary and secondary fluors
    \item Overlap of emission and absorption spectra of solvent and fluors
\end{itemize}
\begin{figure}
    \centering
    \includegraphics[width=0.45\textwidth]{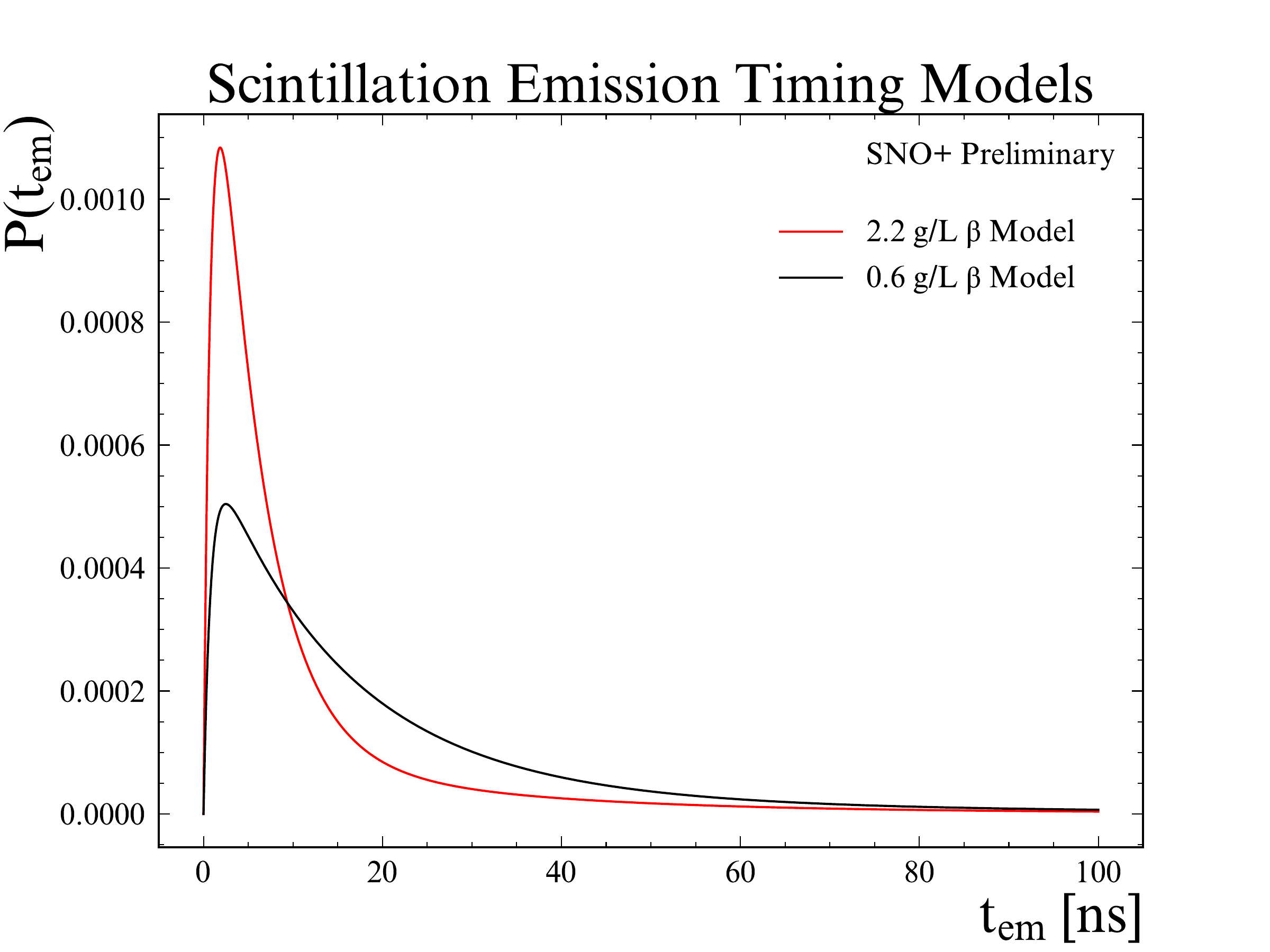}
    \qquad
    \includegraphics[width=0.45\textwidth]{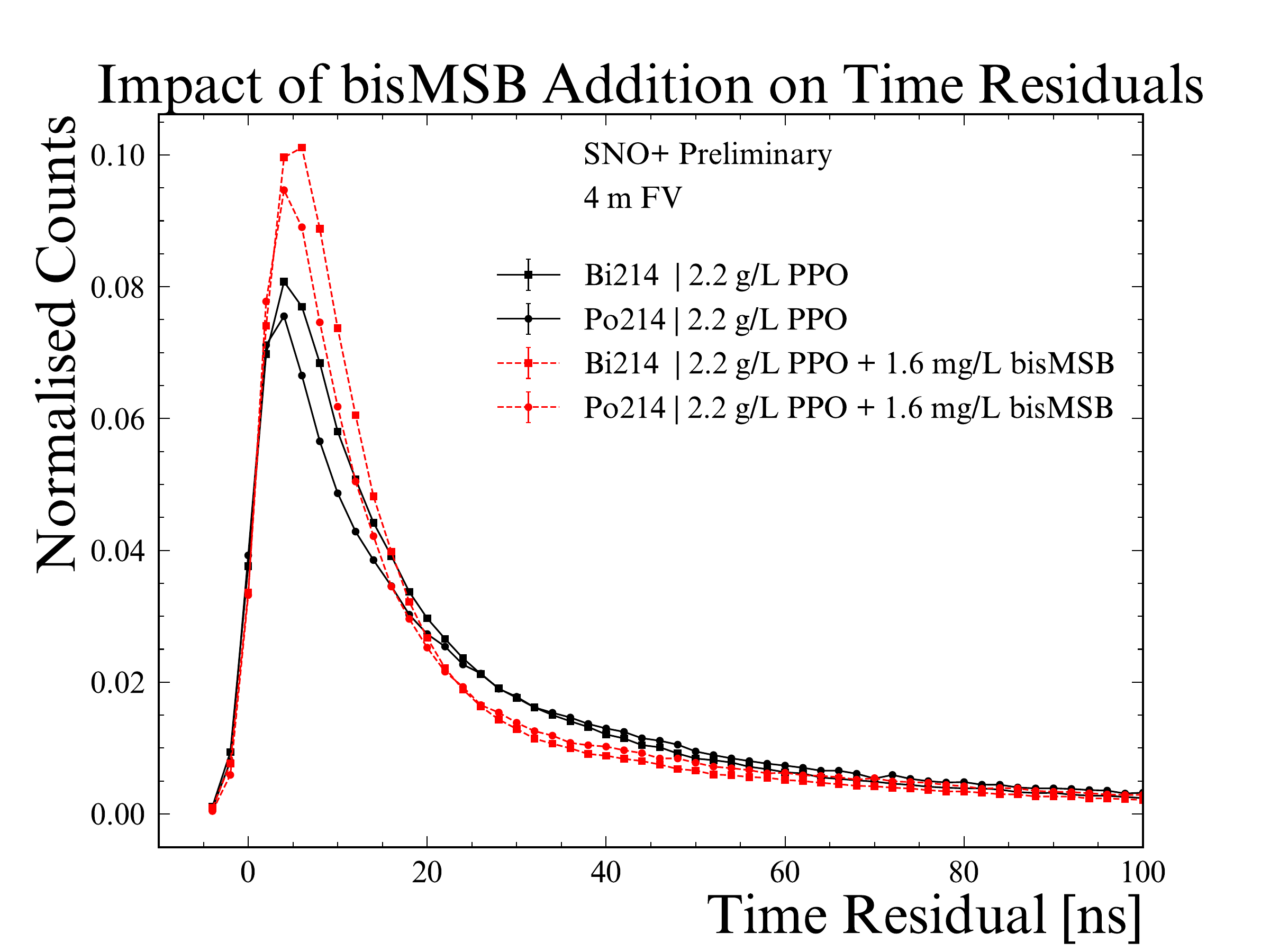}
    \caption{Left: Difference between tuned $\beta^-$ emission time models for Partial-Fill (black, 0.6 g/L PPO) and Full-Fill (red, 2.2 g/L PPO) phase. These models correspond to the intrinsic timing of the scintillator in the presence of $\beta^-$ particles, and are tuned by comparing tagged $^{214}$Bi data and MC time residuals. Right: Difference in data time-response between Full-Fill (black, 2.2 g/L PPO) and intermediate bisMSB loading (red, 2.2 g/L PPO + 1.6 mg/L bisMSB) for $^{214}$Bi and $^{214}$Po.}%
    \label{fig:example}%
\end{figure}
To deal with this complexity, an empirical model is used \cite{Anderson_2021}:
\begin{equation}
    f(t) = \sum_i A_i\frac{e^{-\frac{t}{t_i}}-e^{-\frac{t}{t_r}}}{t_i - t_r}
\end{equation}
This model consists of a sum of exponential terms, with each term characterised by an amplitude, $A_i$, a decay time constant, $t_i$, and a rise time, $t_r$, common to every term. For each operational phase, the empirical model is tuned to match the observed time response obtained from in-situ $^{214}$Bi or $^{214}$Po events.
This tuning has historically been performed via a simple grid search over the model parameters, guided by benchtop measurements. However, due to the large quantities of Monte Carlo produced by the grid search (a data set for every combination of timing parameters tested), investigations into using a more efficient Bayesian optimisation tuning algorithm have been performed. Upcoming calibrations of the bisMSB loaded phase will take advantage of the new algorithm.

\section{BiPo214 as an In-Situ Calibration Source}
BiPo214 events were selected as the in-situ calibration source for three reasons \cite{wang2022a}:
\begin{enumerate}
    \item They provide a source of $\beta^-$ and $\alpha$ interactions, which characterise the majority of charged particle interactions of interest in the detector
    \item Their rate is sufficient to produce samples with good statistics
    \item They occur in coincidence with one another, allowing a clean sample to be extracted from the detector
\end{enumerate}
These coincidences are characterised by a prompt $^{214}$Bi $\beta^-$ decay, followed by a delayed $^{214}$Po $\alpha$ decay. Applying prompt and delayed energy cuts, in addition to an inter-event distance and time cut, allows a clean sample of BiPo214 events to be extracted from the detector, with negligible contamination by accidentals.
This is verified by examining the $\Delta t$ distribution (see figure 2) between the prompt $^{214}$Bi and delayed $^{214}$Po: fitting a decaying exponential should recover the $^{214}$Po half-life of $t_{1/2} = 164.3 $ $\mu s$ for sufficiently pure samples. In addition, distortions in the energy distributions are key indicators of accidental contamination, which are not observed.
\begin{figure}%
    \centering
    \includegraphics[width=0.45\textwidth]{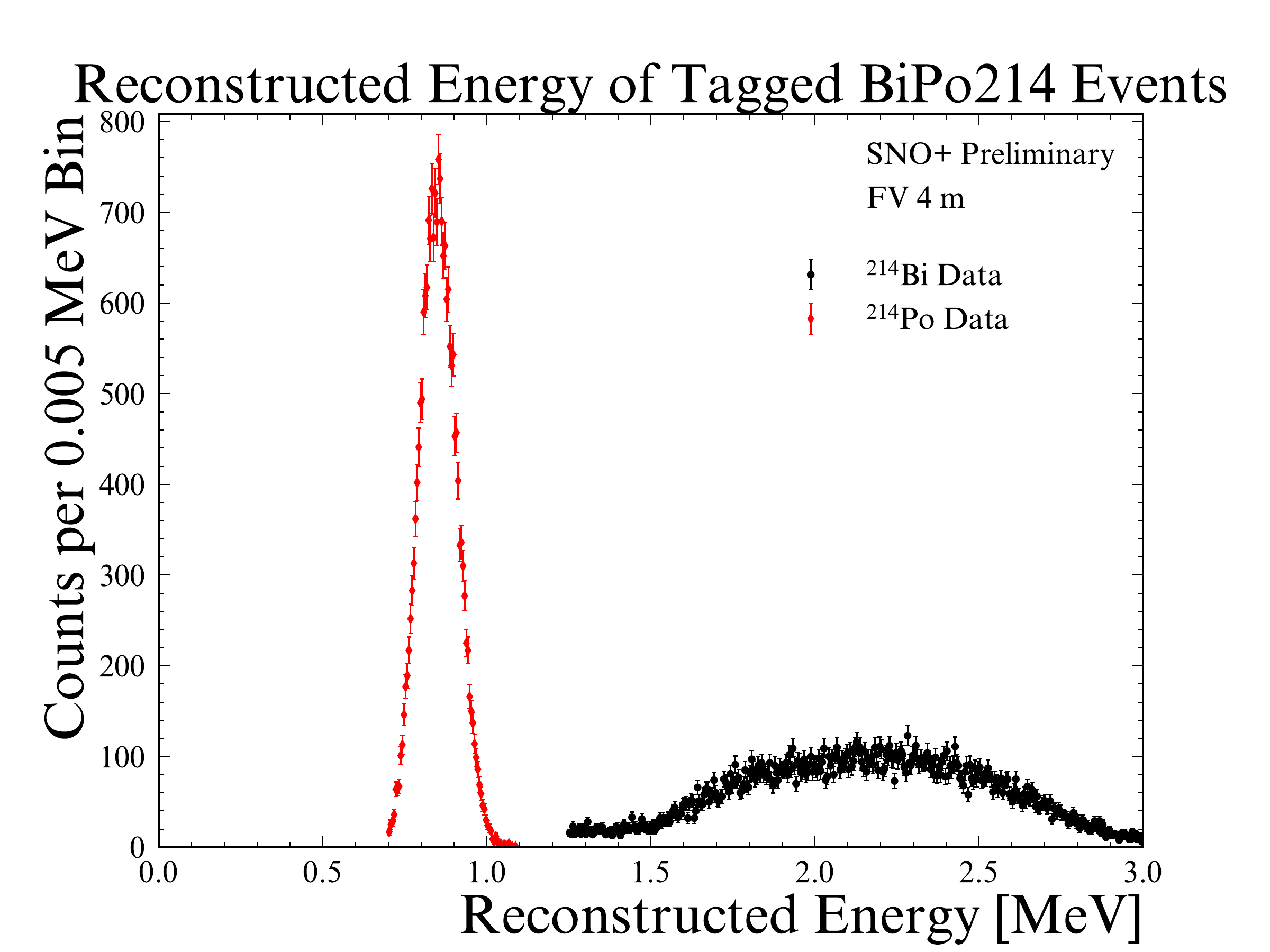}
    \qquad
    \includegraphics[width=0.45\textwidth]{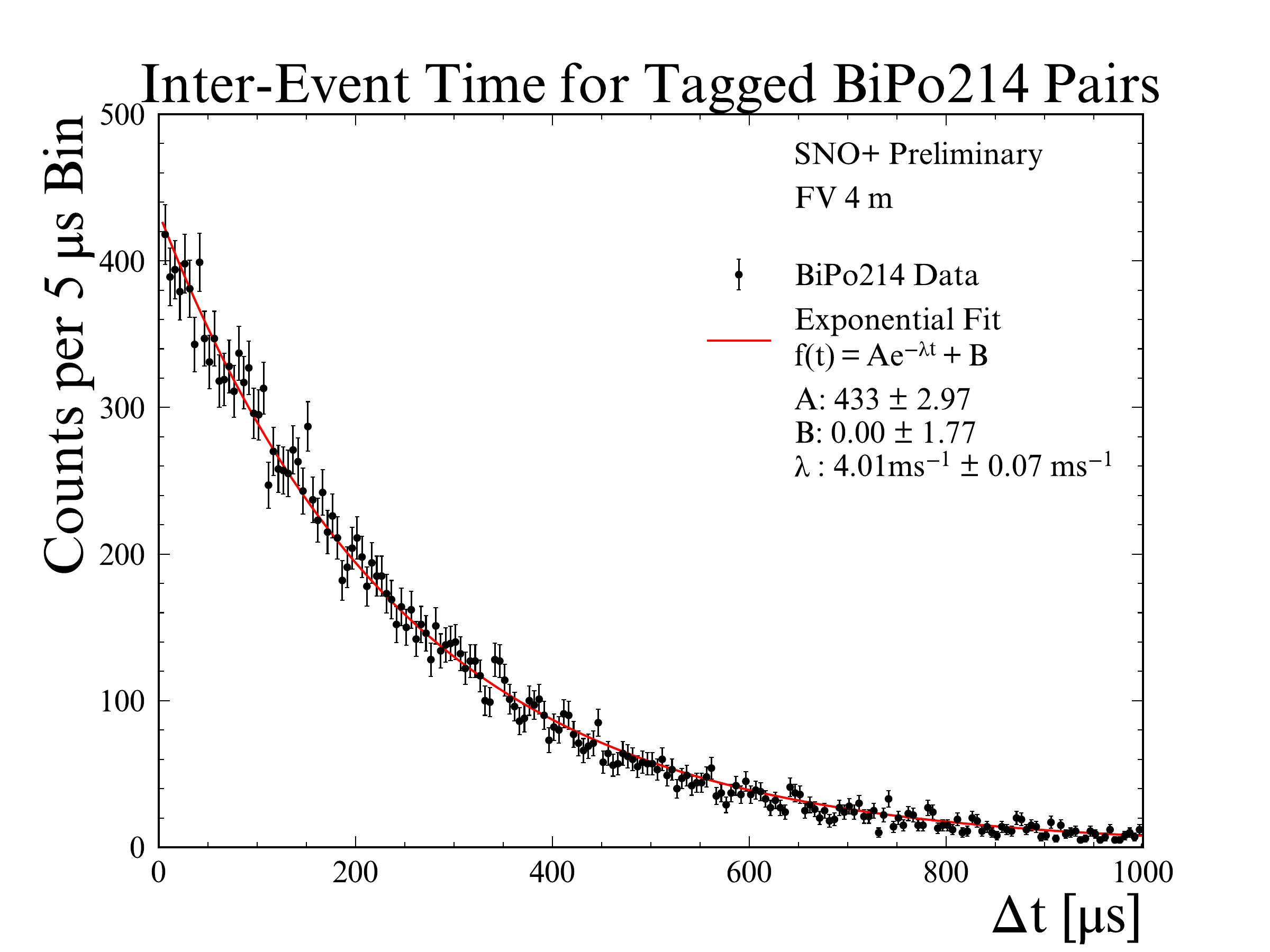}
    \caption{Left: Energy spectra of extracted $^{214}$Bi (black) and $^{214}$Po (red) events. Right: Fitted $\Delta$t distribution.}%
    \label{fig:example}%
\end{figure}

\section{Full-Fill (2.2 g/L PPO) Emission Time Calibration Result}
A set of amplitudes, decay time constants and a rise time was obtained for both $\beta^-$ and $\alpha$ events in the full-fill phase.
The target quantity is a distribution of `time-residuals’: 
\begin{equation}
    t_{residual} = t_{hit} - t_{t.o.f} - t_{ev}
\end{equation}
For every PMT hit time, $t_{hit}$, the time of flight, $t_{t.o.f}$, and the reconstructed event time, $t_{ev}$, is subtracted. The residual time is taken to be the intrinsic emission time distribution of the liquid scintillator, convolved with the detector response. Assuming a good understanding of the detector response, the time residual agreement between data and simulated events is maximised when the model accurately describes the intrinsic emission time profile of the scintillator.

The resulting simulated time residual distributions show good agreement to the data across a broad range of times.
\section{Multi-Site Event Discrimination using Calibrated Full-Fill Emission Time Profiles}
\begin{figure}%
    \centering
    \includegraphics[width=0.45\textwidth]{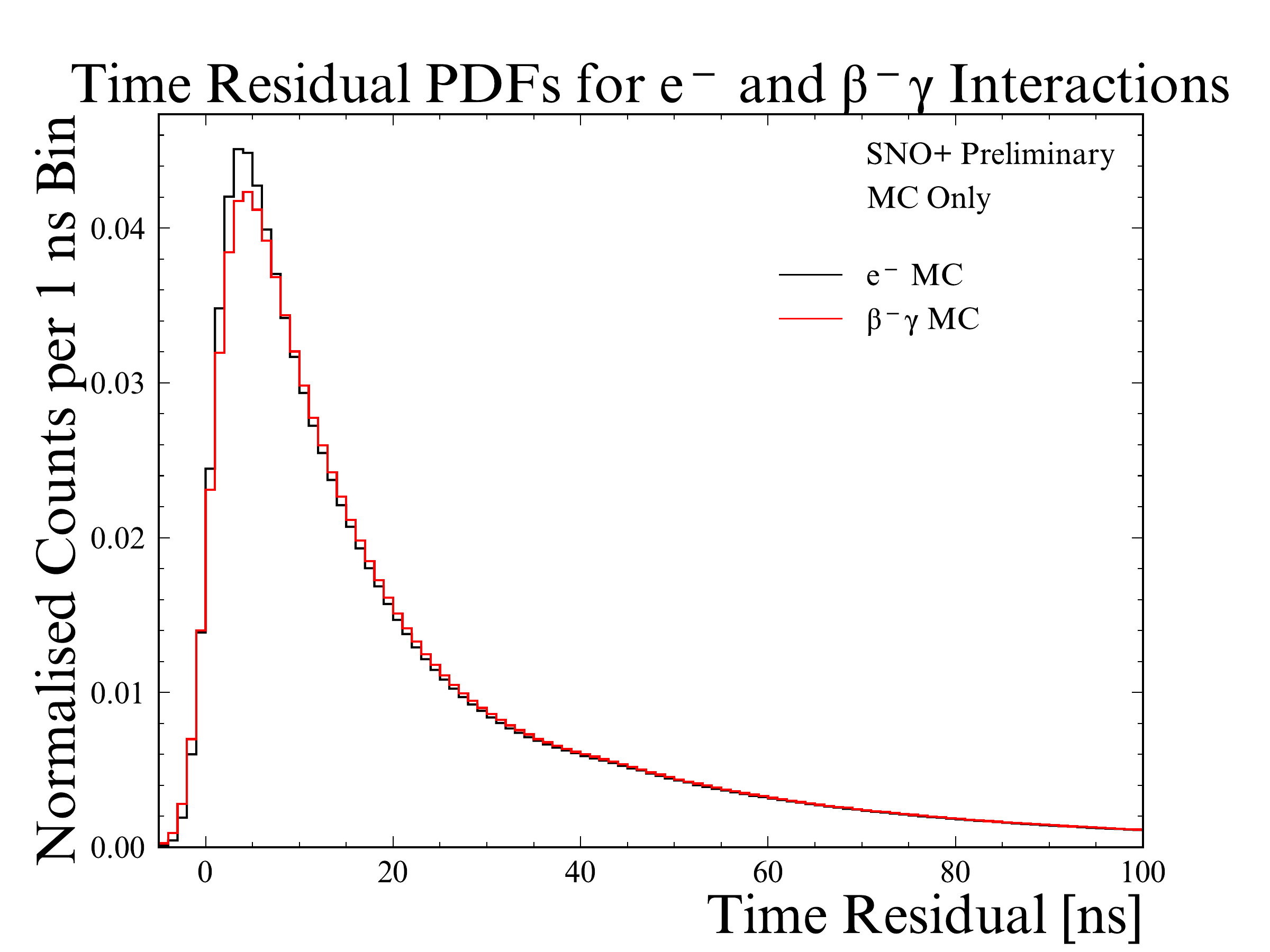}
    \qquad
    \includegraphics[width=0.45\textwidth]{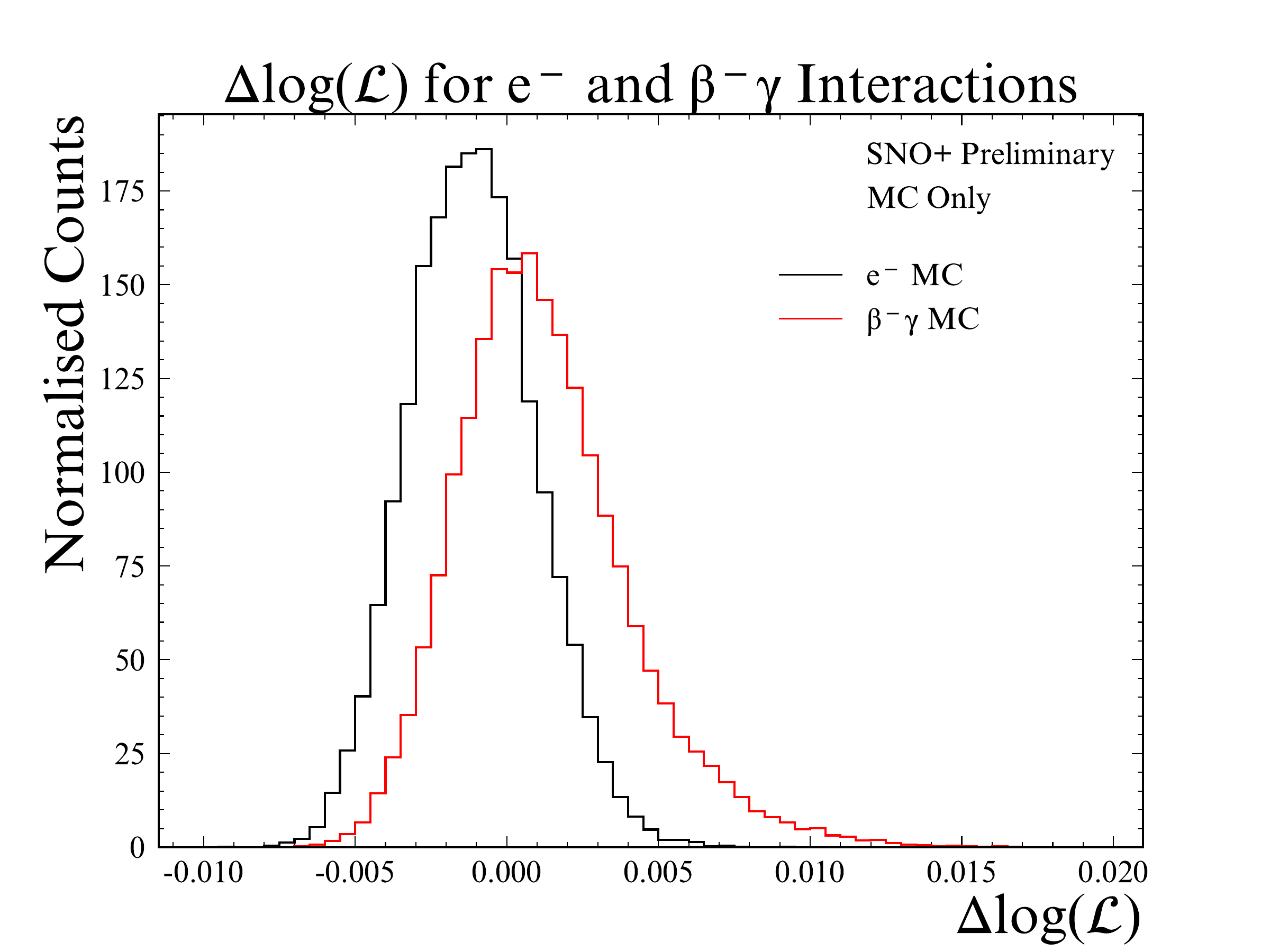}
    \caption{Left: Time residual PDFs used to evaluate the multi-site log-likelihood discriminant for single-site (black) and multi-site events (red). Right: log-likelihood ratio discriminant for single-site (black) and multi-site (red) events.}%
    \label{fig:example}%
\end{figure}
Discrimination between events which deposit energy at single points (single-site events, typically $\beta$-like events) and events which have a diffuse energy deposition (multi-site events, typically $\gamma$ events) is desirable for a number of analyses, including solar neutrino and neutrinoless double-beta studies. In both cases, many of the backgrounds are multi-site and the signal of interest is a single-site interaction. The spread of energy depositions in multi-site events leads to broader time residual distributions as compared to single-point energy depositions. A well calibrated emission time model accurately reproduces this broadening, allowing the creation of well-determined time residual PDFs for each event class (see figure 3). These PDFs are used in conjunction with an event's time residuals to calculate the log-likelihood ratio discriminant \cite{Dunger_2019} \cite{tereza}: 
\begin{equation}
    \Delta log(\mathcal{L}) = \frac{1}{N_{hit}}\sum_i^{N_{hit}} log\left(\frac{P_S(t_{res}^i)}{P_B(t_{res}^i)}\right)
\end{equation}
Where $N_{hit}$ is the total number of PMT hits in an event, $P_S$/$P_B$ are the signal and background time residual PDFs and $t_{res}^i$ is the time residual for the $i^{th}$ hit.

An example application is $^8$B solar neutrino studies, where the dominant background around 3.8 MeV is the multi-site $\beta^- \gamma$ decay of $^{208}$Tl; the solar interaction is single-site. Thus, incorporating a measure of the multi-site character of events (e.g. equation 3) into likelihood-based signal extraction frameworks is highly desirable to improve measurement sensitivities in regions dominated by multi-site backgrounds.

\section{Conclusions and Outlook}
The SNO+ detector has concluded its full-fill PPO addition campaign (2.2 g/L PPO), and the bisMSB loading of the scintillator is in progress. A calibration of the scintillation emission timing model has been performed using in-situ radioactive sources, namely Bismuth-Polonium214 coincidence decays. This BiPo214 source provides pure samples of $\beta^-$ and $\alpha$ events, respectively, whilst the method preserves the radiopurity of the detector. By exploiting the coincident nature of the decays and applying appropriate inter-event cuts, the contamination of the samples by accidentals is negligible.

Using the well-calibrated emission timing model, it is possible to exploit the differences between single-site events (e.g. $^{8}$B solar neutrino $\beta^-$ interactions) and multi-site events (e.g. internal $^{208}$Tl $\beta^- \gamma$ decays). Studies are ongoing into incorporating a log-likelihood ratio based multisite discriminant into the pre-existing solar signal extraction framework, improving measurement sensitivity in multi-site background dominanted regions around 3.8 MeV.

\section*{Acknowledgements}
This work is supported by ASRIP, CIFAR, CFI, DF, DOE, ERC, FCT, FedNor, NSERC,
NSF, Ontario MRI, Queen’s University, STFC, and UC Berkeley, and have benefited
from services provided by EGI, GridPP and Compute Canada. Rafael Hunt-Stokes is supported by the STFC. We thank Vale and SNOLAB for their valuable support.

\printbibliography
% \bibitem{...}

% \end{thebibliography}\endgroup

\end{document}